\documentclass[aps,prb,reprint,groupedaddress]{revtex4-1}
\usepackage[dvipdfmx]{graphicx}
\usepackage{amsmath}
\usepackage{mathcomp}
\usepackage{bm}
\usepackage{color}
\usepackage{longtable}
\begin{document}
%\preprint{}

\title{
%A working note on Sm$_{2}$Fe$_{17}$N$_{x}$
Finite-temperature magnetic properties of Sm$_{2}$Fe$_{17}$N$_{x}$ \\
using an ab-initio effective spin model
}


\author{
$^{1}$Shogo Yamashita, $^{1}$Daiki Suzuki, $^{1}$Takuya Yoshioka, \\
$^{1,2}$Hiroki Tsuchiura, 
and $^{3}$Pavel Nov\'{a}k}
%\thanks{}
\affiliation{
$^1$Department of Applied Physics, Tohoku University, Aoba 6-6-05, Aoba-ku, 
 Sendai 980-8579, Japan, \\
 $^2$Center for Spintronics Research Network, Tohoku University, Sendai 980-8577, Japan \\
 %}
%\affiliation{$^2$Center for Spintronics Research Network, Tohoku University, 
% Sendai 980-8577, Japan}
%\affiliation{
$^3$Institute of Physics, Academy of Sciences of the Czech Republic, Cukrovarnick\'{a} 10, 
162 53 Praha 6, Czech Republic
}

\date{\today}

%=========================================================
\begin{abstract}
%The unusual magnetic anisotropy properties of Sm$_{2}$Fe$_{17}$N$_{x}$ systems have been investigated theoretically and experimentally but their structural causes are not thoroughly understood. 
In this study, we investigate the finite-temperature magnetic properties of Sm$_{2}$Fe$_{17}$N$_{x}$ ($x=0, 3$) using an effective spin model constructed based on the information obtained by first-principles
calculations.
%We find that the model based on the trivalent electronic configuration 
We find that assuming the plausible trivalent Sm$^{3+}$ configuration results in a model 
that can satisfactorily describe the magnetization curves of Sm$_{2}$Fe$_{17}$N$_{3}$.
By contrast, the model based on the divalent Sm$^{2+}$ configuration is suitable to reproduce 
the magnetization curves of Sm$_{2}$Fe$_{17}$.
%the magnetization curves can be well described by the model for $x=3$,
%but for $x=0$, 
These results expand the understanding of how electronic structure affects the magnetic properties of these compounds. 
\end{abstract}
%\keywords{}

\maketitle

%  \subsection{}
%  \subsubsection{}

%%%%%%%%%%%%%%%%%%%%%%%%%%%%%%%%%%%%%%%%%
\section{Introduction}
%%%%%%%%%%%%%%%%%%%%%%%%%%%%%%%%%%%%%%%%%
%%
%\noindent
%1. Sm$_{2}$Fe$_{17}$N$_{x}$の磁気特性は窒化によって大きく変わる．

The Sm$_{2}$Fe$_{17}$N$_{x}$ nitrides have unusual magnetic anisotropy properties.
The system with $x=3$ is a commercially successful permanent magnet and is
well known to exhibit stronger uniaxial magnetocrystalline anisotropy than Nd$_{2}$Fe$_{14}$B.
In contrast, the system with $x=0$, binary Sm$_{2}$Fe$_{17}$, has weak 
planar anisotropy, and a recent experiment has revealed that the magnetization orientation
slightly deviates from the basal plane by about 10$\tcdegree$ at low temperatures \cite{diop}.
Because the iron sublattice of this system is expected to have planar magnetic anisotropy
analogous to Y$_{2}$Fe$_{17}$, this observation indicates that the local magnetic 
anisotropy due to the Sm ions is uniaxial but comparatively weak in Sm$_{2}$Fe$_{17}$.
Thus, the nitrogenation process sensitively changes the electronic states around the Sm ions,
resulting in a sign change of the magnetocrystalline anisotropy.

%%
%\noindent
%2．第一原理計算の先行研究

Several efforts have been made to theoretically clarify the electronic states of 
Sm$_{2}$Fe$_{17}$N$_{x}$ based on first-principles calculations 
\cite{min,steinbeck,nekrasov,pandey,ogura}.
Because modern first-principles calculations still do not treat 4f electrons properly,
some additional treatment such as the so-called open core method, the local  spin density approximation with Hubbard correction (LSDA+$U$), or 
the self-interaction correction (SIC) is needed to evaluate the magnetic anisotropy 
of these systems.
Steinbeck {\it et al.} \cite{steinbeck} calculated the 4f crystal field parameters (CFPs)
for the Sm ions
based on the open core method, assuming a plausible trivalent Sm$^{3+}$ configuration.
They found that the second-order CFP $A_{2}^{0}$, which dominates the local magnetic
anisotropy of Sm, is enhanced in amplitude by nitrogenation.
 Knyazev {\it et al.} \cite{nekrasov} calculated the electronic structure and optical properties of Sm$_{2}$Fe$_{17}$ and Tm$_{2}$Fe$_{17}$ using the LSDA+$U$ method. The calculated total magnetic moment including the orbital correction to the contribution from 4f electrons is  in good agreement with  the experimentally measured value\cite{Kou}. However, they did not discuss the magnetic anisotropy of the systems. Pandey {\it et al.} \cite{pandey} calculated the magnetocrystalline anisotropy energy of Sm$_{2}$Fe$_{17}$ and Sm$_{2}$Fe$_{17}$N$_{3}$. The predicted directions of magnetocrystalline 
anisotropy for Sm$_{2}$Fe$_{17}$ and Sm$_{2}$Fe$_{17}$N$_{3}$ are in good agreement with the experimental observations. However, the calculated total moments are smaller than the experimentally measured values\cite{Buschow,McNeely}.
According to their results, the calculated spin magnetic moments of the Sm ions are close to that of trivalent Sm$^{3+}$.
%According to  their results of the calculated spin magnetic moments, the Sm ions are  close to  trivalent Sm$^{3+}$.}
Recently, Ogura {\it et al.} \cite{ogura} carried out self-consistent Korringa--Kohn--Rostoker coherent potential approximation (KKR-CPA) calculations with a SIC treatment,
and confirmed that the uniaxial magnetocrystalline anisotropy increases with increasing
nitrogen content $x$.
In addition, they claimed that the number of f electrons of each Sm ion
is about 6; that is, a divalent Sm$^{2+}$ configuration is realized, regardless of $x$.
This result is quite intriguing because it has been widely believed that the Sm ions have trivalent 
electronic states, as claimed in several X-ray absorption spectroscopy (XAS) experiments\cite{XAS}.
%Moreover, several x-ray adsorption spectroscopy (XAS) experiments have claimed that
%the Sm ions in Sm$_{2}$Fe$_{17}$N$_{x}$ systems are trivalent.
%
%%
%\noindent
%3. 問題点（XASなど）
%

We notice that these theoretical studies  except the work of Pandey {\it et al.} concluded uniaxial magnetocrystalline 
anisotropy not only for 
Sm$_{2}$Fe$_{17}$N$_{3}$, but also for Sm$_{2}$Fe$_{17}$, which is 
inconsistent with the reported experimental observations. %for Sm$_{2}$Fe$_{17}$.
It should be also noted here that the KKR-CPA with SIC calculations showed that
the uniaxial magnetic anisotropy of Sm$_{2}$Fe$_{17}$ is fairly small.
This is intuitively understandable because the 4f electron clouds of divalent Sm$^{2+}$
should be more spherical than those of trivalent Sm$^{3+}$, resulting in
a weaker uniaxial local magnetic anisotropy. Although information about the electronic states of Sm is quite important to understand
the magnetic properties, in particular, the magnetocrystalline anisotropy, it is difficult to
extend the conclusions of the previous studies by using the first-principles calculation
method itself.
Therefore, we instead focus on the finite-temperature magnetic properties of the systems.
We construct a single-ion model, which has long been used to describe phenomenologically
the magnetic properties of rare-earth based materials, but base it on
first-principles calculations.
The necessary information to construct the model is as follows: the magnetic moments
of each ion, the exchange field acting on the 4f electrons, and the CFPs.
Once we construct the model, including the crystal-field Hamiltonian, we can compute
the magnetic anisotropy due to the rare-earth ions for arbitrary temperatures
in the standard statistical mechanical way.
For this purpose, the most suitable way to treat the 4f electrons in first-principles calculations
is the open core method because we can easily control the electronic structure and valency
of the Sm ions.

In this study, we investigate the magnetization curves of Sm$_{2}$Fe$_{17}$N$_{x}$ $(x=0, 3)$
for several temperatures using the single-ion model based on first-principles calculations.
We prepare the model in two ways: (1) assuming trivalent Sm ions, and (2) assuming divalent Sm ions.
We show the differences between the magnetization curves obtained by (1) and (2),
and discuss which valency and electronic configurations plausibly describe
the experimentally obtained finite-temperature magnetic properties.

%%%%%%%%%%%%%%%%%%%%%%%%%%%%%%%%%%%%%%%%%
\section{Electronic structure calculations and model construction}
%%%%%%%%%%%%%%%%%%%%%%%%%%%%%%%%%%%%%%%%%
%%
%\noindent
%{\color{red} 1. シングルイオンモデル．}

We use the 
%well known 
single-ion Hamiltonian\cite{Wijn,Sankar,Yamada,Richter,Franse} to describe the finite-temperature magnetic
properties of Sm$_{2}$Fe$_{17}$N$_{x}$ $(x=0, 3)$. The Hamiltonian of the $i$-th Sm ion is given as
\begin{align}
 &{\cal H}(i) \nonumber \\
 &= \lambda {\bm L}_{\mathrm{4f}} {\bm \cdot} {\bm S}_{\mathrm{4f}}
 	+ 2 {\bm H}_{\textrm{m}}(T) {\bm \cdot} {\bm S}_{\mathrm{4f}}
	+ {\cal H}_{\mathrm{CEF}}(i)
	+ \left( {\bm L}_{\mathrm{4f}} + 2 {\bm S}_{\mathrm{4f}} \right) {\bm \cdot} {\bm H}_{\mathrm{ex}},
\label{YamadaKato}	
\end{align}
where ${\bm L}_{\mathrm{4f}}$ and ${\bm S}_{\mathrm{4f}}$ are the total spin and total angular momentum operators, $\lambda$ is the spin-orbit coupling constant of the 4f shell, ${\bm H}_{\mathrm{ex}}$ is the  external magnetic field, and
${\bm H}_{\textrm{m}}(T)$ is the exchange mean-field acting on the spin components of 4f electrons.
%%%%%%%%%%%%%%%%%%%%%%%%%%%%%%%%%%%%%%%%%%%
%入れ替え
%%%%%%%%%%%%%%%%%%%%%%%%%%%%%%%%%%%%%%%%%%%
The crystal-field Hamiltonian $H_{\textrm{CEF}}(i)$ is expressed in terms of the tensor operator method as follows\cite{Sankar,Yamada}:
\begin{align}
\label{CFHamil}
&H_{\textrm{CEF}}(i)=  A_2^0(i)\sum_{j}r^2_{j}2U_2^0(\theta_j,\phi_j) \nonumber \\ 
&+A_4^0(i)\sum_{j}r^4_{j}8U_{4}^0(\theta_j,\phi_j)  
+A_6^0(i)\sum_{j}r_{j}^616U_{6}^0(\theta_j,\phi_j)\nonumber  \\ \
&+A_{6}^6(i)\sum_{j}r_j^6 16\left(\frac{1}{231}\right)^\frac{1}{2}\left[U_6^6(\theta_j,\phi_j) 
+U_{6}^{-6}(\theta_j,\phi_j)\right],
\end{align}
where $A_{l}^{m}(i)$ are the CFPs at the $i$-th rare-earth ion site.
%The crystal field potential is expanded by the tesseral harmonics function $Z_{l,m}(\theta,\phi)$.
In this work, assuming that all sites are equivalent,  we neglect contributions from $A_{4}^3(i)$ and $A_{6}^3(i)$. The tensor operator $U_{l}^{m}$ is given by
\begin{align}
U_{l}^{m}(\theta_j,\phi_j)=\sqrt{\frac{4\pi}{2l+1}}Y_{l,m}(\theta_j,\phi_j),
\end{align}
where $Y_{l,m}$ is the spherical harmonics function. 
Matrix elements of $\sum_{j}r^l_jU_{l}^m(\theta_j,\phi_j)$ are expressed using 3-j and 6-j symbols as follows\cite{Sankar,Yamada}:
\begin{align}
\langle J,J_z,L,S|\sum_{j}r^l_{j}U_{l}^{m}(\theta_j,\phi_j)|J',J'_z,L,S\rangle \nonumber  \\ 
 =(-1)^{(L+S-J_z+J+J')}\sqrt{(2J+1)(2J'+1)} \nonumber
 \\
\times  \begin{pmatrix}
J & J' & l \\
-J_z & J'_z & m \\
\end{pmatrix}
\begin{Bmatrix}
L & L & l \\
J & J' & S \\
\end{Bmatrix}
\langle L||U_l||L \rangle \langle r^l \rangle,
\end{align}
where $\langle L||U_l||L \rangle$ is the appropriate set of reduced matrix elements given in Table \ref{reduced}.
\begin{table}
 \caption{\label{reduced} The reduced matrix elements \cite{Yamada} used in our calculations.}
\begin{ruledtabular}
 \begin{tabular}{cccccccccccc}
 	     &   $\langle L||U_2||L \rangle$ &  $\langle L||U_4||L \rangle$ & $\langle L||U_6||L \rangle$ \\ \hline
%	\multirow{2}{*}{Sm$_{2}$Fe$_{17}$}
	 Sm$^{3+}$ & $\frac{1}{3}\left(\frac{2\cdot11\cdot13}{15}\right)^\frac{1}{2}$  & $\frac{2}{3}(\frac{2\cdot 13}{11})^{\frac{1}{2}}$ & $-10(\frac{5\cdot 17}{3\cdot 11 \cdot 13})^{\frac{1}{2}}$\\ \hline
					     Sm$^{2+}$\footnote{For $\textrm{Sm}^{2+}$ we show the values of $\textrm{Eu}^{3+}$.} & $2(\frac{7}{15})^{\frac{1}{2}}$& $-(\frac{14}{11})^{\frac{1}{2}}$&   $10(\frac{7}{3\cdot11\cdot13})^{\frac{1}{2}}$\\ 

 \end{tabular}
\end{ruledtabular}
\end{table}
We note that $J$ multiplets are taken into account up to the fifth excited state in this calculation.
We use $L=5$ and $S=\frac{5}{2}$ for Sm$^{3+}$, and $L=3$ and $S=3$ for Sm$^{2+}$.
 %

%We also obtain $\langle r^{2}\rangle = 1.02a_{0}$ for the Nd $4f$-density, \\ \
Once we have set the Hamiltonian for the $i$-th Sm ion, we can obtain the free energy for the 4f partial system based on the 
statistical mechanical procedure, as 
\begin{equation}
F(i) = -k_{\rm B}T \ln {\rm Tr} \exp 
	\left[ -\frac{ {\cal H}(i) } { k_{\rm B}T } \right].
%\label{EA}
\end{equation}
%
%\begin{align}
%F(i)=-\frac{1}{\beta} \log \sum_{s=1}^{\sum_{J} 2J+1} \textrm{e}^{-\beta{E_s(i)}}
%\end{align}
%where we assume that the direction of the magnetic moment of the Nd ion is parallel
%to that of $\left| {\bm H}_{m} \right| $.

%We calculate expectation value of Magnetic moment from eigen vectors of this Hamiltonian.
%\begin{align}
%{\cal H}_{\textrm{R}}(i)|i,s\rangle=E_s(i)|i,s\rangle \\ 
%\ | i,s \rangle=\sum_{j_z}a^s_{J,J_z}| J,J_z\rangle
%\end{align}
%Also, the crystal field Hamiltonian ${\cal H}_{\mathrm{CEF}}$ is defined as
%\begin{equation}
 %{\cal H}_{\mathrm{CEF}} = \sum_{l,m}\Theta_{l}A_{l}^{m}\langle r^{l}\rangle\hat{O}_{l}^{m} ,
 %\label{CFHamil}
%\end{equation}
%where $\hat{O}_{l}^{m}$ are the Stevens {\it operator equivalents}, 
%$A_{l}^{m}$ are the crystal field coefficients, and  $\Theta_{l}$ are
%the reduced matrix elements \cite{Stevens,Hutchings}.
%{\color{red} 1.5 歴史配置に困っている} \\
This single-ion Hamiltonian has a quite long history, and in the early days of the study 
of rare-earth-based permanent magnets, the model parameters such as $|{\bm H}_{\textrm{m}}(0)|$ and $A_{l}^{m}$
were determined by multi-parameter fitting calculations to the experimental magnetization
curves of single crystals \cite{Yamada}.
Now, we can determine these parameters based on the information of the electronic states
of the systems using first-principles calculations.
Recently, we have confirmed that this procedure successfully describes the observed 
magnetization curves and the temperature dependences of anisotropy constants 
of {\it R}$_{2}$Fe$_{14}$B ({\it R} = Dy, Ho)\cite{Yoshiokasan} and SmFe$_{12}$\cite{Yoshioka1-12}.

%%
%\noindent
%{\color{red} 2. 第一原理計算的パラメータ見積もり法} \\
%\ 
We use the first-principles calculations method to determine the CFPs. Once the electronic state calculations are completed, we can compute the CFPs
$A_{l}^{m} \langle r^{l} \rangle$ in Eq. (\ref{CFHamil}) based on the well-known formula \cite{Novak,Divis1,Divis2}
\begin{equation}
 A_{l}^{m}\langle r^{l}\rangle = a_{lm} \int_{0}^{R_{\mathrm{MT}}}dr r^{2}|R_{\rm 4f}(r)|^{2}V_{l}^{m}(r) ,
%\nonumber \\
%                                        &&              + \left. \int_{R_{MT}}^{\infty}dr r^{2}|R_{\rm 4f}(r)|^{2}W_{L}^{M} \right) ,            
\label{alm}
\end{equation}
%
%
%\begin{equation}
 %\langle r^{l}\rangle = \int_{0}^{R_{\mathrm{MT}}}dr r^{l+2}|R_{\rm 4f}(r)|^{2} ,
%\label{rlmean}
%\end{equation}
%
where $V_{l}^{m}(r)$ is the component of the partial wave expansion of the total Coulomb 
potential of the rare-earth ions
within the atomic sphere of radius $R_{\mathrm{MT}}$.
 $a_{lm}$ are numerical factors, specifically, $a_{20}=\frac{1}{4}\sqrt{\frac{5}{\pi}}$, $a_{40}=\frac{3}{16}\sqrt{\frac{1}{\pi}}$, $a_{60}=\frac{1}{32}\sqrt{\frac{13}{\pi}}$, $a_{66}=\frac{231}{64}\sqrt{\frac{26}{231\pi}}$, and $R_{\textrm{4f}}(r)$ is the radial shape of the localized 4f charge density of 
the rare-earth ions. We can directly obtain $V_{l}^{m}(r)$ from the density functional theory (DFT) potential calculated by WIEN2k\cite{wien2k}.
Moreover, to simulate the localized 4f electronic states in the system, we use the {\it classical} open core
method, in which we switch off the hybridization between 4f and valence 5d and 6p states
and treat the 4f states in the spherical part of the potential as atomic-like core states \cite{Novak}.
Thus, the function $R_{\textrm{4f}}(r)$ in Eq. (\ref{alm}) can be obtained by performing separate
atomic calculations of the electronic structure of an isolated rare-earth atom.
The details of the calculations are provided in previous studies \cite{Novak,Divis1,Divis2}.

% \noindent
% {\color{red} 3. 磁気特性計算法} \\
Using the polar coordinates $(\theta,\phi)$, which are the zenith and azimuth angles defined with the $c$-axis as the $z$-axis, the total free energy of the system is given by
\begin{align}
F(\mbox{\boldmath $H$}_{\textrm{ex}},{H}_{\textrm{m}}(T),T,\theta,\phi)&=\sum_{i=1}^{N_{\textrm R}} F(i)+N_{\textrm{Fe}}K_{\textrm{Fe}}(T)\sin^2\theta  \nonumber \\ 
&-\mbox{\boldmath ${M}$}_{\textrm{Fe}}(T)\cdot \mbox {\boldmath $H$}_{\mathrm{ex}},
\label{EA}
\end{align}
where $N_\textrm{R}$ and $N_\textrm{Fe}$ are the numbers of Sm ions and Fe ions, respectively; ${H}_{\textrm{m}}(T)=|\mbox{\boldmath $H$}_{\textrm{m}}(T)|$; and
$K_{\textrm{Fe}}(T)$ is the anisotropy constant of Fe per atom given in Table \ref{AnisoFe}.  The same notation is used as in Eq. \eqref{YamadaKato}.
\begin{table} 
 \caption{\label{AnisoFe} The anisotropy constants $K_{\textrm {Fe}}$ [K]  per single Fe atom. ${\cite{diop,Inami,Brennan}}$}
\begin{ruledtabular}
 \begin{tabular}{cccccccc}
 	    &  $4.2 \textrm{K}$ & $300 \textrm{K}$ \\ \hline
%	\multirow{2}{*}{Sm$_{2}$Fe$_{17}$}
	 $\textrm{Sm}_{2}\textrm{Fe}_{17}$ & $-4.14$  & $-0.45$  \\ \hline
	 $\textrm{Sm}_{2}\textrm{Fe}_{17}\textrm{N}_{3}$ & $-2.60$& $-1.30$  \\ 

 \end{tabular}
\end{ruledtabular}
\end{table}
The temperature dependence of $\bm{M}_{\textrm{Fe}}(T)$ is described by the Kuz'min formula as follows\cite{Kuzmin}:
\begin{align}
\mbox{\boldmath${M_{\textrm{Fe}}}$}(T)=\mbox{\boldmath${M_{\textrm{Fe}}}$}(0)\left[1-s\left( \frac{T}{T_{\textrm{c}}}\right)^{\frac{3}{2}}-(1-s)\left(\frac{T}{T_{\textrm{c}}}\right)^p\right]^\frac{1}{3}
\end{align}
where $\bm{M}_{\textrm{Fe}}(0)$ is the total magnetic moment of the system except the magnetic moment of the Sm ions. The calculated results of $\bm{M}_{\textrm{Fe}}(0)$ are summarized in Table \ref{magmom}. 
 The temperature dependence of  ${\bm H}_{\textrm{m}}(T)$ is also described by the Kuz'min formula\cite{Kuzmin}:
\begin{align}
{\bm H}_{\textrm{m}}(T)={\bm H}_{\textrm{m}}(0)\left[1-s\left( \frac{T}{T_\textrm{c}}\right)^{\frac{3}{2}}-(1-s)\left(\frac{T}{T_\textrm{c}}\right)^p\right]^\frac{1}{3}.
\end{align}
In this system, we use $s=0.7$ and $p=5/2$ given by Kuz'min \cite{Kuzmin}. The Curie temperatures $T_\textrm{c}$ of $\textrm{Sm}_2\textrm{Fe}_{17}$ and $\textrm{Sm}_2\textrm{Fe}_{17}\textrm{N}_3$ are 380K\cite{diop} and 752K\cite{koyama}, respectively.
Calculating Eq. (\ref{EA}) for given ($\theta$, $\phi$), we obtain the angular dependence of the total free energy.
We note that the directions of $\bm H_{\textrm{m}}$ $(T)$ and $\bm M_{\textrm{Fe}}$ $(T)$ are anti-parallel.
The equilibrium directions of $\mbox{\boldmath${M_{\textrm{Fe}}}$}(T)$ and $\bm H_{\textrm{m}}$$(T)$ are evaluated using the minimum point of the total free energy of the system. We numerically examine all the angles $(\theta,\phi)$ that correlate with the energy minimum point. These angles determine the direction of the Fe sublattice magnetization in equilibrium.
The finite-temperature magnetic moment of the $i$-th Sm ion $\mbox{\boldmath${M_{\textrm{R}}}$}^i(T)$ is given by
\begin{align}
\mbox{\boldmath${M_{\textrm{R}}}$}^i(T)&= -\sum_{\textrm{s}}\langle i,\textrm{s} |\mbox{\boldmath $L$}_{\mathrm{4f}}+2\mbox{\boldmath $S$}_{\mathrm{4f}}| i,\textrm{s}\rangle \textrm{e}^{-{E_{\textrm{s}}(i)/k_\textrm{B}T}}/Z(i), 
\end{align}
\begin{align}
Z(i)=\sum_{\textrm{s}}\textrm{e}^{-\frac{E_{\textrm{s}}(i)}{k_{\textrm{B}}T}},
\end{align}
where $E_{\textrm{s}}(i)$ and $|i,\textrm{s} \rangle$ are an eigenvalue and eigenvector, respectively, of the following equation:
 \begin{align}
{\cal H} (i)|i,\textrm{s}\rangle=E_{\textrm{s}}(i)|i,\textrm{s}\rangle.
\end{align}
The total magnetic moment of the system $\mbox{\boldmath${M}$}(T)$ is represented by
\begin{align}
\mbox{\boldmath${M}$}(T)=\sum_{i}\mbox{\boldmath${M_{\textrm{R}}}$}^i(T)+\mbox{\boldmath${M_{\textrm{Fe}}}$}(T).
\label{Totalmom}
\end{align}
We plot the magnetization curves of both compounds at 4.2K and 300K because the anisotropy constants of Fe summarized in Table \ref{AnisoFe} are measured at a specific temperature.
In order to calculate the magnetic anisotropy constants $K_1(T)$ and $K_2(T)$, we assume the following expansion:
\begin{align}
F(\mbox{\boldmath $H$}_{\textrm{ex}},{ H}_{\textrm{m}}(T),T,\theta,\phi)-N_{\textrm{Fe}}K_{\textrm{Fe}}(T)\sin^2\theta \nonumber
 \\ =K_1(T)\sin^2\theta+K_2(T)\sin^4\theta+\cdots.
\end{align}
We use Maclaurin's expansion to obtain the anisotropy constants as follows\cite{Sasaki}:
\begin{align}
K_1(T)=\frac{1}{2}\frac{{\partial}^2F(0,\mbox { $H_{\textrm{m}}$}(T),T,\theta,\phi)}{ {\partial}\theta^2}|_{\theta=0},
\end{align}
\begin{align}
K_2(T)=\frac{1}{3}K_1(T)+\frac{1}{24}\frac{{\partial}^4 F(0,\mbox { $H_{\textrm{m}}$}(T),T,\theta,\phi) }{{\partial} \theta^4}|_{\theta=0}.
\end{align}
 We note that the magnetic anisotropy of Fe is not included in $K_1(T)$ and $K_2(T)$ in this work, because we cannot treat the temperature dependence of the magnetic anisotropy of Fe theoretically.
{\color{blue} 
%サイト毎の特徴はどうか．●
%また，3価と2価ではどうか．●
%Smに近いFeが影響を受けているはず．●
%Steinbeckの結果と比較してどうか？●
%total  momentsは，Smの寄与を含む．
%ただし，Smの磁気モーメントについては，軌道の寄与が含まれるため，スピンモデルを用いて見積もる．
}
%---------------------------------------------------------------------------------------------
%	Table I, spin magnetic moments
%---------------------------------------------------------------------------------------------

%---------------------------------------------------------------------------------------------

%%
%%\noindent
%% and $\langle r^{2}\rangle = 0.727a_{0}$ for Dy,
%where $a_{0}$ is the Bohr radius.
%The computational detail is given in ref. \cite{tsuchi}.
%
We calculate $| \boldmath {H}_{\textrm{m}}$$(0)|$ in accordance with the method of Brooks $et \ al.$\cite{Brooks}.
%{\color{blue} We also need to describe here how we can estimate $| H_{m} |$. 
%交換磁場はどうやって計算する？●
%Also, it is necessary to note that, for Sm ions, we take account of the excited states 
%of the 4f-multiplets. 
%It may be nice to show here the eigenvalues.
%}
%
We summarize the model parameters, including CFPs, $H_{\textrm{m}}(0)$, and $\lambda$,
in Table \ref{CFPs}.

%---------------------------------------------------------------------------------------------
%---------------------------------------------------------------------------------------------

%%

\begin{figure*}[htb]
\centering
\includegraphics[clip,width=17cm]{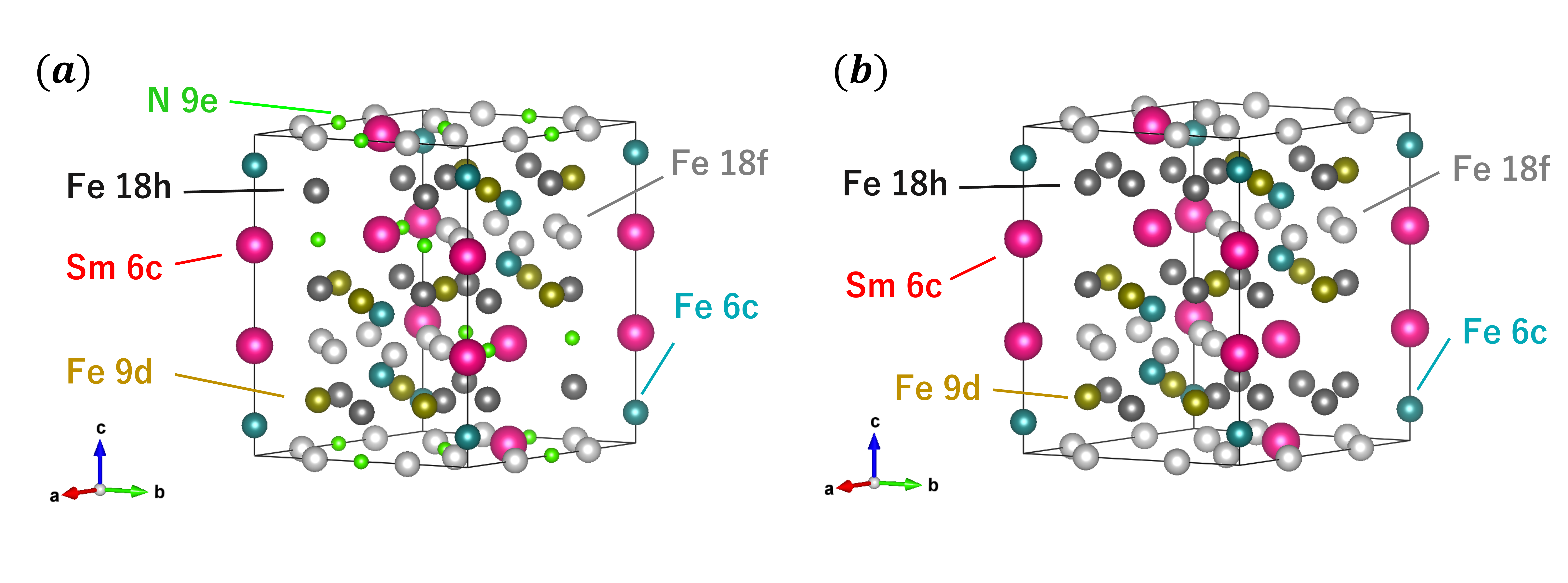}
\caption{ Crystal structures of the hexagonal unit cell and non-equivalent atomic positions of  (a)  Sm$_{2}$Fe$_{17}$N$_{3}$ \cite{Inami} and (b) Sm$_{2}$Fe$_{17}$.\cite{Teresiak}
}
%\caption{\label{label}Figure caption}
\label{struct}
\end{figure*}

%{\color{red} 4. パラメータ詳細} \\ \
For the computation of the necessary parameters for the single-ion model, %Sm$_{2}$Fe$_{17}$N$_{x}$,
we use the WIEN2k code (Version 16.1), adopting the generalized-gradient approximation form for
the exchange-correlation functional.
Here, the lattice constants of the unit cell are set to the experimental values of
$a = b = 8.557${\AA}   and $c = 12.448${\AA}  for Sm$_{2}$Fe$_{17}$\cite{Teresiak}, and
$a = b = 8.7206${\AA}  and $c = 12.63445${\AA}  for Sm$_{2}$Fe$_{17}$N$_{3}$\cite{Inami}.
The space group of both compounds is $R\bar{3}m$ (No. 166).
The atomic sphere radii are taken to be 3.0, 1.96, and 1.6 a.u. for Sm, Fe, and N,
respectively. Spin-orbit coupling is not considered in our first-principles calculations.
The number of basis functions of Sm$_{2}$Fe$_{17}$ and Sm$_{2}$Fe$_{17}$N$_{3}$ are taken to be 
1609 and 2934, respectively, and $4\times4\times4$ \  $k$ points are sampled in the Brillouin zone. {$RK_{\textrm{MAX}}$ is taken to be 7.0 in our calculations.} 
{Crystal structure and non-equivalent atomic positions of  Sm$_{2}$Fe$_{17}$N$_{3}$ and Sm$_{2}$Fe$_{17}$ are shown in FIG. \ref{struct}.}
%-----------------------------------------------------------------------------------------------

%---------------------------------------------------
%	Fig. 1: 
%---------------------------------------------------
%\begin{figure}[htb]
%\begin{center}
%\includegraphics[height=5cm]{DyNdFeB.eps}
%\includegraphics[clip,width=7cm]{levels.pdf}
%\end{center}
%
%\caption{
%The eigenvalues of the $J$-multiplet of the Sm ion in the effective spin model.
%{\color{blue} Sm$_{2}$Fe$_{17}$N$_{x}$の場合の図が必要．}
%}
%\caption{\label{label}Figure caption}
%\label{Sm_levels}
%\end{figure}
%---------------------------------------------------
%{\color{blue}
%Smの$J$多重項の固有値分布を示す．
%交換磁場が強いため，低温でも励起$J$を考慮する必要があることに言及し，山田-加藤を引用する．
%}

%{\color{blue}
%Smの磁気モーメントについて，基底Jのみの場合と励起状態を取り入れた場合について触れておく．
%励起Jを取り入れること，つまり量子論的取り扱いをすることが，磁化を議論する上で
%重要であることを指摘する．
%}

%{\color{blue}
%図\ref{Sm_levels}を提示した後に，J多重項を考慮することは，Smの異方性だけでなく，
%磁気モーメントを考える上でも不可欠であることについて述べる．
%the ground Jのみの場合と，励起状態を取り入れた場合の磁気モーメントを比較する．
%}

%%%%%%%%%%%%%%%%%%%%%%%%%%%%%%%%%%%%%%%%%
\section{Results and Discussion}
%%%%%%%%%%%%%%%%%%%%%%%%%%%%%%%%%%%%%%%%%
%
\subsection{Electronic structure}
We explain the calculated electronic structure for $\textrm{Sm}_2\textrm{Fe}_{17}$ and $\textrm{Sm}_2\textrm{Fe}_{17}\textrm{N}_{3}$.
We show the spin magnetic moments of the Fe ions on each site, the total spin magnetic moments in the cell except the contribution from the Sm spin magnetic moments, and the total magnetic moments in the unit cell at 0K in Table \ref{magmom},
for comparison with the results of previous studies\cite{steinbeck,ogura}.
The total magnetic moments in the unit cell are calculated by Eq. \eqref{YamadaKato} and Eq. \eqref{Totalmom}, because the magnetic moments of the Sm ions contain the contribution from the orbital magnetic moments of 4f orbitals. 
The nearest neighbor sites for the Sm ions in  $\textrm{Sm}_2\textrm{Fe}_{17}$ are 18f sites.
In contrast, the nearest neighbor sites for the Sm ions in $\textrm{Sm}_2\textrm{Fe}_{17}\textrm{N}_{3}$  are 18h sites\cite{steinbeck,Inami}.
The difference of the valency does not affect the spin magnetic moments of each site drastically.
We find that the nitrogenation causes the enhancement of the magnetic moments at 9d sites and reduction of those at 18f sites in both valency. Similar effects are discussed in Ogura $et \ al.$  However, no enhancement of magnetic moments at 6c sites was found in this work. 
We note that the magnetic moments of each Fe site in trivalent $\textrm{Sm}_2\textrm{Fe}_{17}$ are quite different from the results of  Steinbeck $et \ al$., especially on 9d sites, and the magnetic moments of each Fe site in trivalent $\textrm{Sm}_2\textrm{Fe}_{17}\textrm{N}_{3}$ are similar to their results except that of 6c sites\cite{steinbeck}. {For the trivalent configuration, we can see that  the total magnetic moments of Sm ions in $\textrm{Sm}_2\textrm{Fe}_{17}$ and $\textrm{Sm}_2\textrm{Fe}_{17}\textrm{N}_{3}$ are $0.48 \mu_{\textrm{B}}$ and $0.41 \mu_{\textrm{B}}$, respectively. In contrast, for the divalent configuration, we can see the negative contribution; the total magnetic moments of Sm ions are $-2.70 \mu_{\textrm{B}}$ and $-2.68 \mu_{\textrm{B}}$ for $\textrm{Sm}_2\textrm{Fe}_{17}$ and $\textrm{Sm}_2\textrm{Fe}_{17}\textrm{N}_{3}$, respectively. The comparison of the calculated total magnetic moment at low temperature with experimentally measured values are discussed in sections I\hspace{-.1em}I\hspace{-.1em}I. B and I\hspace{-.1em}I\hspace{-.1em}I. C.} 
Next, we show the CFPs $A_{l}^{m} \langle r^{l} \rangle$  calculated by Eq. (\ref{alm}) in Table \ref{CFPs}.
We can see that the CFPs $A_{l}^{m} \langle r^{l} \rangle$ of $\textrm{Sm}_2\textrm{Fe}_{17}\textrm{N}_{3}$ do not change drastically, except $A_{2}^{0} \langle r^{2} \rangle$, when we change the valency of the Sm ions. We also note that the values of  $A_{2}^{0} \langle r^{2} \rangle$ for $\textrm{Sm}_2\textrm{Fe}_{17}\textrm{N}_{3}$ are the largest regardless of the its valency. This implies that $\textrm{Sm}_2\textrm{Fe}_{17}\textrm{N}_{3}$ shows uniaxial anisotropy, regardless of the valency of the Sm ions.
However, we can see that the CFP $A_{2}^{0} \langle r^{2} \rangle$ of  $\textrm{Sm}_2\textrm{Fe}_{17}$ is quite different in trivalent and divalent results. $A_{2}^{0} \langle r^{2} \rangle$ in the trivalent result is the largest among the CFPs; however, $A_{2}^{0} \langle r^{2} \rangle$ in the divalent result is quite small compared with the others. This implies that the difference of the valency of the Sm ions causes the change of the anisotropy of the system.
 \begin{table}%[H] %add [H] placement to break table across pages
 %\caption{\label{magmom} The spin magnetic moments  {\color{red} in unit $\mu_{\textrm{B}}$ }within the atomic spheres
 %of each Fe ion, the total spin magnetic moment in the cell except the contribution from the Sm spin magnetic moments, and the 
% total magnetic moment per unit cell calculated by our effective single-ion model for each valency. The results of the spin magnetic moments of each Fe site of previous works\cite{steinbeck,ogura} are also shown. Steinbeck $et \ al.$ treated the Sm ions as trivalent in a similar method with the open core method. Ogura $et \ al.$ did not use the open core method, thus the valency of the Sm ions is not shown. We extracted the values of magnetic moments from Fig. 2 of Ogura $et \ al.$\cite{ogura}
% } 
\caption{\label{magmom}Spin magnetic moments   in unit $\mu_{\textrm{B}}$ of each Fe ion, total spin magnetic moment in the cell except the contribution from the Sm, and
 total magnetic moment per unit cell calculated by our effective single-ion model for each valency, with comparison to previous works\cite{steinbeck,ogura}.}
 \begin{ruledtabular}
 \begin{tabular}{cccccccc}
 	     & Valency  & 6c & 9d & 18f & 18h  & $\bm M_{\textrm{Fe}}(0)$ & Total \\ \hline
%	\multirow{2}{*}{Sm$_{2}$Fe$_{17}$}
	Present work\\ Sm$_{2}$Fe$_{17}$  & Sm$^{3+}$ & 2.67 & 2.21 & 2.49 & 2.39   &39.72& 40.19\\ 
					    & Sm$^{2+}$ & 2.71 & 2.20 & 2.50 & 2.42  & 41.06 &  35.66\\ \hline
	Present work\\ Sm$_{2}$Fe$_{17}$N$_{3}$  
					    & Sm$^{3+}$ & 2.65 & 2.48 & 2.18 & {2.36} &38.80  & 39.41 \\ 
					    & Sm$^{2+}$ & 2.69 & 2.49 & { 2.10} & 2.40  & 38.96& 33.61\\  \hline
					     Steinbeck $et \ al.$\cite{steinbeck} \footnote{The Sm ions are treated as trivalent in a similar method with the open core method. } \\ Sm$_{2}$Fe$_{17}$  &Sm$^{3+}$&2.41&1.57&2.22&2.02&--&--\\ Sm$_{2}$Fe$_{17}$N$_{3}$ &Sm$^{3+}$&2.36&2.45&2.13&2.42&--&--\\   \hline
					     Ogura $et \ al.$\cite{ogura} \footnote{They did not use the open core method, thus the valency of the Sm ions is not shown. We extracted the values of magnetic moments from Fig. 2 of Ogura $et \ al.$\cite{ogura} } \\ Sm$_{2}$Fe$_{17}$  &--&2.53&2.15&2.42&2.17&--&--\\ Sm$_{2}$Fe$_{17}$N$_{3}$
					     &--&2.92&2.84&2.14&2.12&--&--\\
					    
% 		& valency & 18h & 18f & 9d & 6c & total
%	Sm$_{2}$Fe$_{17}$ & Sm$^{3+}$ & 2.20 & 2.47 & 2.17 & 2.50 & 35.2 \\
%	Sm$_{2}$Fe$_{17}$ & Sm$^{2+}$ & 2.19 & 2.49 & 2.18 & 2.55 & 35.1 \\
\label{moment}
 \end{tabular}
\end{ruledtabular}
\end{table}
%\begin{longtable}
\begin{table*}%[H] %add [H] placement to break table across pages
 \caption{\label{CFPs} The crystal field parameters {$A_{l}^{m} \langle r^{l} \rangle $ [K]} and the amplitude of the 
 exchange mean-field $| \boldmath {H}_{\textrm{m}}$$(0)|$ {[K]} computed using first-principles calculations, and the spin-orbit coupling constants $\lambda$ {[K]} calculated from the experimental results\cite{McClure}.}
\begin{ruledtabular}
%\scalebox{0.5}[0.9]{
\small
\begin{tabular*}{\columnwidth}{ccccccccc}
 	   & Valency  &$A_{2}^{0} \langle r^{2} \rangle $
	   		      &$A_{4}^{0} \langle r^{4} \rangle $
			      &$A_{6}^{0} \langle r^{6} \rangle $
			      &$A_{6}^{6} \langle r^{6} \rangle $
			      & $|\boldmath {H}_{\textrm{m}}$$(0)|$ & $\lambda$& \\ \hline
%	\multirow{2}{*}{Sm$_{2}$Fe$_{17}$}
	Sm$_{2}$Fe$_{17}$  & Sm$^{3+}$ & -176.1 & -18.27 & -0.122 & 43.07 & 354.1 & 350&\\ 
					    & Sm$^{2+}$ & -8.07 & -17.27 & 0.08 & 50.25 & 372.5 & 387&\\ \hline
	Sm$_{2}$Fe$_{17}$N$_{3}$  & Sm$^{3+}$ & -576.7 & 7.055 & -1.213 & 29.85 & 176.3 & 350&\\ 
	
					    & Sm$^{2+}$ & -718.5 & 11.27 & -1.142 & 28.54 & 150.5 &  387&\\ 
% 		& valency & 18h & 18f & 9d & 6c & total
%	Sm$_{2}$Fe$_{17}$ & Sm$^{3+}$ & 2.20 & 2.47 & 2.17 & 2.50 & 35.2 \\
%	Sm$_{2}$Fe$_{17}$ & Sm$^{2+}$ & 2.19 & 2.49 & 2.18 & 2.55 & 35.1 \\
 \end{tabular*}
 %}
\end{ruledtabular}

\end{table*}
%\end{longtable}
\subsection{Magnetization curves of Sm$_{2}$Fe$_{17}$N$_{3}$ }

\begin{flushleft}
\begin{figure*}[htb]
\includegraphics[clip,width=15cm]{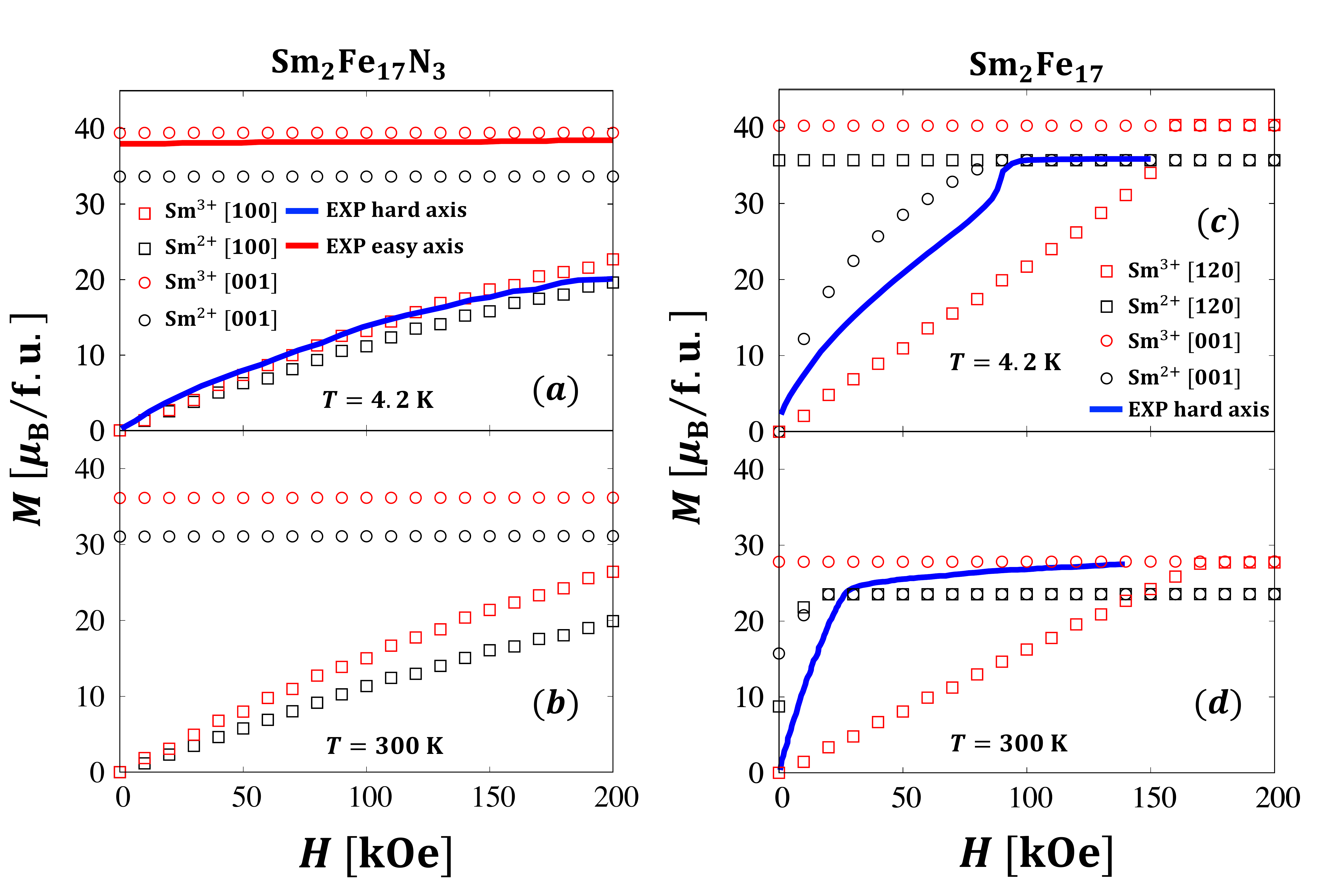}
\caption{The calculated magnetization curves of Sm$_2$Fe$_{17}$N$_3$ at (a) 4.2K and (b) 300K and Sm$_2$Fe$_{17}$ at (c) 4.2K and (d) 300K. Sm$^{2+}$ and Sm$^{3+}$ are indicated by black and red plots, respectively, where $H$ [kOe] is the strength of the applied external magnetic field.
In (a) and (b), the circles and squares indicate the results along the principal axis of [001] and [100] direction, respectively, and
in (c) and (d), the circles and squares indicate the results along the principal axis of [001] and [120] direction, respectively.
The experimental curves shown by red and blue lines in (a) at 4.2K are extracted from Koyama $et \ al$.\cite{koyama},
in which the curves for 300K are not shown. 
 The experimental curves in (c) and (d) at 4.2K and 300K are extracted from Diop $et \  al$. \cite{diop}
} 
%\caption{\label{label}Figure caption}
\label{curve}
\end{figure*}
\end{flushleft}

%--------------------------
% Sm2Fe17Nx (x=3, T=4.2K)
%--------------------------
First, we look at the magnetization curves of Sm$_{2}$Fe$_{17}$N$_{3}$ calculated by the effective spin model.
%because the previous first-principles calculations all successfully described
%its strong experimentally observed uniaxial magnetic anisotropy.
%
FIG. \ref{curve} (a) shows the magnetization curves along the [001] 
and [100] directions at $T=4.2$K, obtained by using the effective spin model
with  the trivalent Sm$^{3+}$ (red) and the divalent Sm$^{2+}$ (black). {The experimental results for Sm$_{2}$Fe$_{17}$N$_{3.1}$ reported by Koyama $et \ al.$\cite{koyama} are indicated by solid curves in FIG. \ref{curve} (a).}
We can clearly see that the [001] direction is the easy axis of this system,
regardless of the valency of the Sm ions.
However, as we noted {in the section I\hspace{-.1em}I\hspace{-.1em}I. A}, the saturation magnetizations $M_{\textrm{s}}$ {at low temperature} for Sm$^{3+}$
and Sm$^{2+}$ are clearly different.
{We also notice here that the results for the trivalent model in FIG. \ref{curve} (a)  satisfactorily reproduce
the experimentally observed magnetization curves. %{\color{red} (ここは削除？for Sm$_{2}$Fe$_{17}$N$_{3.1}$ reported in Fig. 13 
%of the paper of Koyama $et \ al.$ \cite{koyama}) }
Thus, 
we can conclude that the Sm$^{3+}$ electronic configuration is realized in
Sm$_{2}$Fe$_{17}$N$_{3}$ compounds.}
%
%---------------------------------------------------
%	Fig. 1: 
%---------------------------------------------------
%\begin{figure}[htb]
%\begin{center}
%\includegraphics[height=5cm]{DyNdFeB.eps}
%\includegraphics[clip,width=7cm]{x3T42.pdf}
%\end{center}
%
%\caption{
% The calculated magnetization curves of Sm$_{2}$Fe$_{17}$N$_{3}$ at $T=4.2$K
%with (a) Sm$^{3+}$ or (b) Sm$^{2+}$ electronic configuration.
%}
%\caption{\label{label}Figure caption}
%\label{MT_x3T42}
%\end{figure}
%--------------------------------------------------
%--------------------------
% x=3, T=300K
%--------------------------
 The magnetization curves of Sm$_{2}$Fe$_{17}$N$_{3}$ at $T=300$K
are shown in FIG. \ref{curve} (b), in the same manner as FIG. \ref{curve} (a).
The saturation magnetization is reduced as changing the valency from Sm$^{3+}$ to  Sm$^{2+}$; however, the qualitative behavior of the curves does not change from FIG. \ref{curve} (a).
%{\color{blue}
%The temperature dependences of the magnetization $M_{s}$ and the magnetic anisotropy
%$H_{A}$ can be mentioned here.●
%}
%
%---------------------------------------------------
%	Fig. 2: 
%---------------------------------------------------
%\begin{figure}[htb]
%\begin{center}
%\includegraphics[height=6cm]{SmFeN300K.pdf}
%\includegraphics[clip,width=7cm]{x3T300.pdf}
%\end{center}
%\caption{
 %The calculated magnetization curves of Sm$_{2}$Fe$_{17}$N$_{3}$ along the principle axes at $T=300$K
 %with  Sm$^{3+}$ (red) or  Sm$^{2+}$ (black) electronic configuration.
 %The circles and squares indicate the results of the [001] and [100] direction, respectively.
 %The squares indicate the results of the [100] direction.
 %{\color{red}丸と四角を述べる}
%}
%\caption{\label{label}Figure caption}
%\label{MT_x3T300}
%\end{figure}
%%---------------------------------------------------

%==================================
\subsection{Magnetization curves of Sm$_{2}$Fe$_{17}$ }
%==================================
Next, we look at the magnetic properties of Sm$_{2}$Fe$_{17}$
described by the effective spin model.
FIG. \ref{curve} (c) shows the calculated magnetization curves along the hard and easy
directions at $T = 4.2$K.
%At first glance, they look similar to those in Fig. \ref{MT_x3T42}, but we notice that 
We can clearly see that the magnetic anisotropy is qualitatively different between the systems
with the trivalent Sm$^{3+}$ and the divalent Sm$^{2+}$ ions.
%{\color{blue}
%3価と2価で，定性的に振る舞いが異なる．
%}
When we assume the divalent Sm$^{2+}$ configuration, the system shows  planar anisotropy.
Thus, the divalent results can reproduce the behavior that is experimentally observed{\cite{diop},}  {as shown by a blue line in FIG. \ref{curve} (c) } at 4.2K.
We also show the curves at 300K in FIG. \ref{curve} (d). The qualitative behavior is almost the same as the results at 4.2K.
It is noted that we can see the spin reorientation phenomenon.
As stated in the Introduction, the orientation of the magnetization of this compound 
was experimentally found to deviate from the basal $a$-$b$ plane by as much as 
10$\tcdegree$ at low temperatures \cite{diop}.
Moreover, at room temperature, the spontaneous magnetization lies within 
the basal plane itself.
%Recalling that the Fe sublattice of this system can be expected to have planar magnetocrystalline
%anisotropy, these observations indicate that the Sm ions have a weak uniaxial magnetic anisotropy 
%in Sm$_{2}$Fe$_{17}$ compounds, causing the small deviation of the magnetization direction 
%from the basal $a$-$b$ plane. 
In our results, for the divalent configuration, when the external magnetic field is not applied, we cannot see the finite magnetization along the [001] direction at 4.2K; however, we can see the finite magnetization along the [001] direction at 300K.
The divalent configuration results can qualitatively reproduce the magnetization curves for Sm$_{2}$Fe$_{17}$, but they cannot explain direction of the spontaneous magnetization at zero field. {From our results, the curves assuming the divalent Sm$^{2+}$ configuration can reproduce the experimental behaviors. However, as stated in the introduction,  XAS experiments have shown that the Sm ions in Sm$_{2}$Fe$_{17}$N$_{x}$ are trivalent Sm$^{3+}$. One possible reason for this discrepancy is the CFPs. The open core method is used in our calculations. The Sm ions are assumed to be in an atomic-like state in this method. The hybridization of 4f and other orbitals is not taken into account in our CFPs.  One possible method beyond the open core method is the Wannierization proposed by Nov\'{a}k {\it et al.\cite{Pavel_Wannier} } In this method, 4f electrons are treated as valence electrons and projected using localized Wannier functions. Thus, we can incorporate the hybridization of 4f and other orbitals in this method. If the hybridization is taken into account, the CFPs and the anisotropy might be changed.}

\subsection{Anisotropy constants $K_{1}(T)$ and $K_{2}(T)$}

%{\color{blue}
%$K_{1}(T)$と$K_{2}(T)$についても触れておくと良いかも．式だけ書いた●
%}

\begin{figure}[htb]
\centering
\includegraphics[clip,width=8cm]{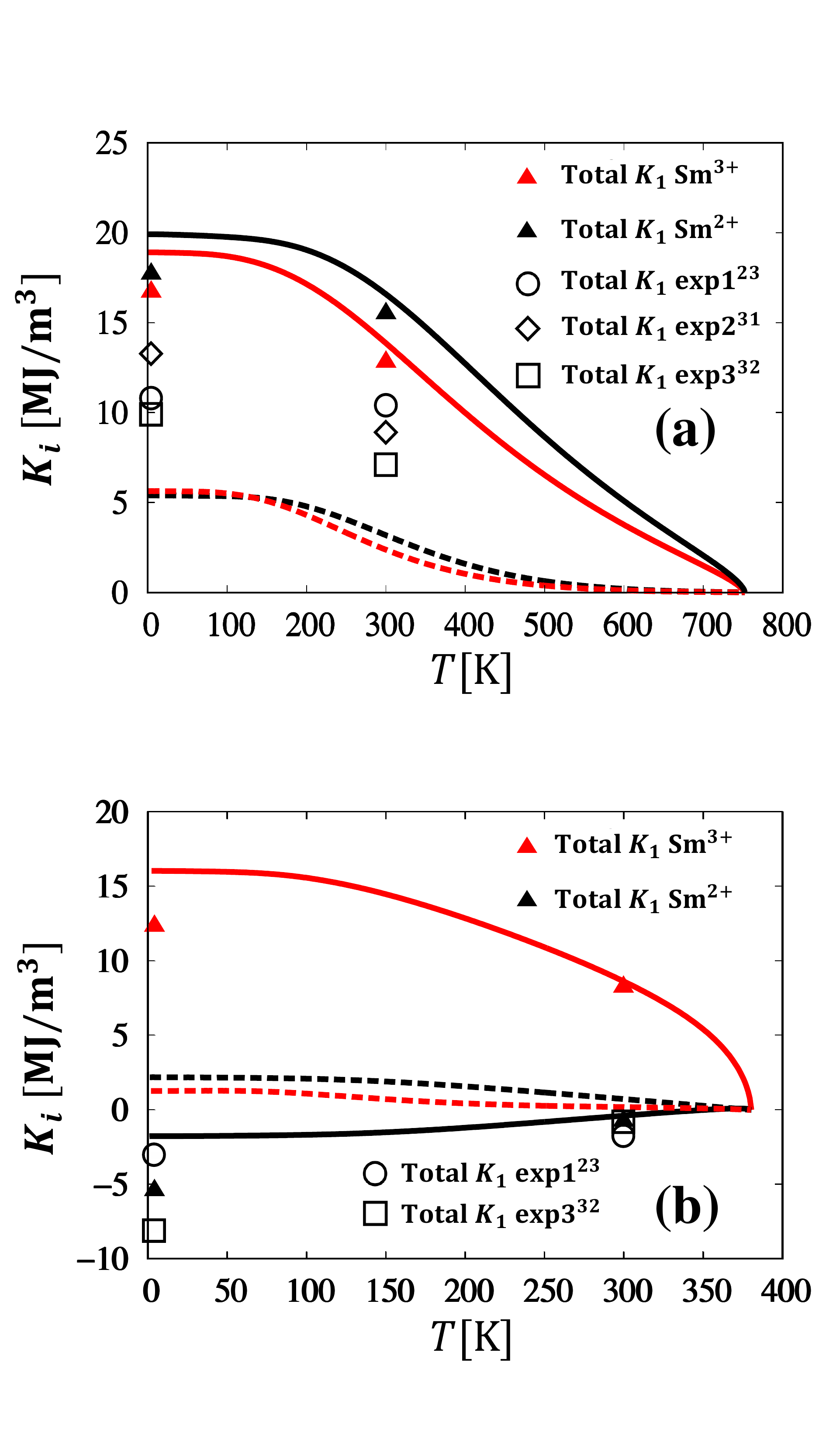}
\caption{
 The calculated magnetic anisotropy constants $K_1$ (solid lines) and $K_2$ (dashed lines) per two Sm ions in (a) Sm$_{2}$Fe$_{17}$N$_{3}$ and (b) Sm$_{2}$Fe$_{17}$
 with Sm$^{3+}$ (red lines) or Sm$^{2+}$ (black lines) electronic configuration. {The total anisotropy constants $K_{1}$ with Fe sublattice contribution $K_{\textrm{Fe}}$   are indicated by red (Sm$^{3+}$) and black (Sm$^{2+}$) triangles. The open shapes indicate the experimentally measured values of $K_{1}$; the open circles correspond to the values of Brennan ${et \ al}$.\cite{Brennan}, the open squares correspond to the values of Isnard ${et \ al}$.\cite{Isnard}, and
the open diamonds  correspond to the values of Wirth ${et\ al}$.\cite{Wirth} }}
%\caption{\label{label}Figure caption}
\label{K1K2}
\end{figure}

We show the temperature dependence of the magnetic anisotropy constants $K_1(T)$ and $K_2(T)$ of Sm$_{2}$Fe$_{17}$N$_{3}$ and Sm$_{2}$Fe$_{17}$ in FIG. \ref{K1K2} {(a) and (b), respectively}.
% (a) and (b). In these figures, $K_1(T)$ and $K_2(T)$ contain the contribution from only two Sm ions in the unit cell. 
 {The anisotropy constants $K_1(T)$ and $K_2(T)$ of Sm$_{2}$Fe$_{17}$N$_{3}$ shown in FIG. \ref{K1K2} (a)} are always positive regardless of the valency of the Sm ions. {The total anisotropy constants $K_1(T)$ which include $K_{\textrm{Fe}}$,  at 4.2K and 300K for each valency are indicated by triangles. The total anisotropy constants $K_1$($T$) are positive at 4.2K and 300K and close to the values measured by Wirth $et \ al.$\cite{Wirth} at low temperature.} This implies that Sm$_{2}$Fe$_{17}$N$_{3}$ shows uniaxial anisotropy at any temperature. The temperature dependence of $K_1(T)$ and $K_2(T)$ of Sm$_{2}$Fe$_{17}$ is shown in FIG. \ref{K1K2} (b) . We can see that the calculation results with trivalent Sm$^{3+}$ show positive $K_1(T)$ and $K_2(T)$ at any temperature. Therefore, Sm$_{2}$Fe$_{17}$ with trivalent Sm$^{3+}$ would show uniaxial anisotropy. 
In contrast, we can see negative $K_1(T)$ in the results with divalent Sm$^{2+}$ configuration. {The total anisotropy constants are also shown in the same manner as Sm$_{2}$Fe$_{17}$N$_{3}$  in FIG. \ref{K1K2} (a). %The total anisotropy constant for Sm$^{2+}$ is much larger than the calculated $K_{1}(T)$ at low temperature.
Our total anisotropy constant for Sm$^{2+}$ is close to the experimental values shown by  Brennan $et \ al.\cite{Brennan}$ and Isnard $et\ al.\cite{Isnard}$ at low temperature. At 300K, our result is in good agreement with the value measured by Isnard $et\  al.$\cite{Isnard}} {, where $K_{1}$ is $-1.75$ {$[\textrm{MJ}/\textrm{m}^3]$} at 300K. Our total anisotropy constants $K_{1}$ for Sm$^{2+}$ are  negative at 4.2K and 300K.} This implies that Sm$_{2}$Fe$_{17}$ with divalent Sm$^{2+}$ would show planar anisotropy at low temperature.
% {\color {red} However, the total anisotropy constant at 300K is not so different from the calculated $K_{1}(T)$.}
%As temperature increase, $K_1(T)$ and $K_2(T)$ become almost same values with opposite sign. This behavior might cause the weak uniaxial magnetic anisotropy.
From the calculated anisotropy constants, the experimentally measured magnetization curve for Sm$_{2}$Fe$_{17}$N$_{3}$ can be explained by both results; however, the curve for Sm$_{2}$Fe$_{17}$ can be explained only by the divalent results.

%\begin{figure}[htb]
%\includegraphics[height=5cm]{DyNdFeB.eps}
%\centering
%\includegraphics[clip,width=18cm]{curve.pdf}

%\caption{}
%\caption{\label{label}Figure caption}
%\label{curve}
%\end{figure}

%\begin{figure}[htb]
%\begin{center}
%\includegraphics[height=5cm]{DyNdFeB.eps}
%\includegraphics[clip,width=7cm]{Ms20191223.pdf}
%\end{center}
%
%\caption{The temperature dependence of $\bm M_s$. (a) shows the one of $\textrm{Sm}_2\textrm{Fe}_{17}$. (b) shows the one of $\textrm{Sm}_2\textrm{Fe}_{17}\textrm{N}_3$.  $\textrm{Sm}^{3+}$(solid lines) $\textrm{Sm}^{2+}$(dashed lines)
%}
%\caption{\label{label}Figure caption}
%\label{Dep_Ms}
%\end{figure}
%%---------------------------------------------------
%\begin{figure}
%\begin{center}
%%\includegraphics[height=5cm]{DyNdFeB.eps}
%\includegraphics[clip,width=8cm]{ph_diag.pdf}
%%\includegraphics[clip,height=4cm]{test2.pdf}
%%\includegraphics[clip,height=4cm]{test2.pdf}
%\end{center}
%%
%\caption{
%The superconducting transition temperatures $T_{c}$ for $\varDelta_{\mathrm{dp}} = 5.1$eV (the red symbols)
%where the charge transfer gap $\varDelta = U_{\mathrm{d}} - \varDelta_{\mathrm{dp}}$ in Fig.\ref{Ueff} is small.
%The results for $\varDelta_{\mathrm{dp}} = 3.9$eV is also shown (blue symbols).
%}
%%\caption{\label{label}Figure caption}
%\label{Tc_diag}
%\end{figure}
%%---------------------------------------------------
%%%%%%%%%%%%%%%%%%%%%%%%%%%%%%%%%%%%%%%%%
\section{Summary}
In this work, we have calculated the magnetization curves and the temperature dependence of the anisotropy constants of Sm$_{2}$Fe$_{17}$ and Sm$_{2}$Fe$_{17}$N$_{3}$ by using the effective spin model based on first-principles calculations with the open core method.
We showed that for Sm$_{2}$Fe$_{17}$N$_{3}$ the curves generated by the Sm$^{3+}$ model are in good agreement with the experimentally measured curves.
{The total anisotropy constants $K_{1}$($T$) calculated for Sm$_{2}$Fe$_{17}$N$_{3}$ assuming Sm$^{3+}$ and Sm$^{2+}$ qualitatively reproduce the experimentally measured behavior, with the results from Sm$^{3+}$  close to the experimental value.}
In contrast, the results for Sm$_{2}$Fe$_{17}$ assuming Sm$^{2+}$  are consistent with the experimentally measured curves. {In addition, the temperature dependence of the total anisotropy constants for Sm$^{2+}$ is consistent with experimentally observed behavior.} However, previously reported XAS experiments\cite{XAS} have found that the Sm ions in Sm$_{2}$Fe$_{17}$N$_{x}$ are trivalent configuration regardless of the nitrogen content $x$.\\
 It is noted that the effective spin model based on DFT calculations with the open core method might not be able to describe the magnetic properties of Sm$_{2}$Fe$_{17}$. One possible reason for this disagreement is that the hybridization effects between the 4f orbitals of  the Sm ions and other orbitals cannot be included in the CFPs by using the open core method. {In the open core method, 4f orbitals are treated as atomic-like core states and the valence effects of them are completely neglected. Taking the valence effects into account might change the CFPs and the magnetic anisotropy of the systems.} For more precise  investigation, further research should apply a more realistic method, such as the Wannierization method, \cite{Pavel_Wannier,yoshi_Wannier} {that can incorporate the hybridization effects of 4f electrons and other orbitals} to evaluate the CFPs.
\begin{acknowledgments}
%One of the authors H. T thanks Kenta Takagi and Wataru Yamaguchi for helpful discussion.
%H. T. and T. Y. would like to thank Satoshi Hirosawa for constant encouragement and 
%constructive suggestions.
%
%{\color {red} This work was supported by JPS KAKENHI Grant Numbers JP18K04678.}
This work was partially supported by ESICMM Grant Number 12016013, and ESICMM is funded
by the Ministry of Education, Culture, Sports, Science and Technology (MEXT).
{S. Y. acknowledges support from GP-Spin at Tohoku University, Japan.}
{ Work of T. Y. was supported by JPS KAKENHI Grant Number JP18K04678.}
Work of P. N. was supported by project SOLID21.
%This work was supported by ESICMM project.
Part of the numerical computations were carried out at the Cyberscience Center, Tohoku University, Japan.
\end{acknowledgments}
%=====================================================

%\section{test}
%\begin{figure}[htb]
%\includegraphics[height=5cm]{DyNdFeB.eps}
%\includegraphics[clip,width=14cm]{SmFeN3.pdf}
%
%\caption{
% The calculated magnetization curves of Sm$_{2}$Fe$_{17}$N$_{3}$ at $T=4.2$K and 300K
% with  Sm$^{3+}$ (Red) or  Sm$^{2+}$ (Black) electronic configuration.{\color{red}丸と四角を述べる}
%}
%\caption{\label{label}Figure caption}
%\label{MT_x3T42}
%\end{figure}


\begin{thebibliography}{9}
%1
\bibitem{diop}
 L. V. B. Diop, M. D. Kuz'min, K. P. Skokov, D. Yu. Karpenkov, and O. Gutfleisch,
 Phys. Rev. B {\bf 94}, 144413 (2016).
%2
\bibitem{min}
 B. I. Min, J.-S. Kang, J. H. Hong, S. W. Jung, J. I. Jeong, Y. P. Lee, S. D. Choi, W. Y. Lee, 
 C. J. Yang, and C. G. Olson,
 J. Phys.: Condens. Matter {\bf 5}, 6911 (1993).
%3
\bibitem{steinbeck}
 L. Steinbeck, M. Richter, U. Nitzsche, and H. Eschrig,
 Phys. Rev. B {\bf 53}, 7111 (1996).
%4
\bibitem{nekrasov}
 Yu. V. Knyazev, Yu. I. Kuz'min, A. G. Kuchin, A. V. Lukoyanov, and I. A. Nekrasov,
 J. Phys.: Condens. Matter {\bf 19}, 116215 (2007).
%5

\bibitem{pandey}
{ T. Pandey, M.-H. Du, and D. S. Parker, Phys. Rev. Appl. {\bf 9}, 034002 (2018).}
\bibitem{Buschow}
{ K. Buschow, Rep. Prog. Phys.  {\bf 54}, 1123 (1991).}
\bibitem{McNeely}
{ D. McNeely and H. Oesterrreicher, J. Less-Common Met. {\bf 44}, 183 (1976). }

\bibitem{ogura}
 M. Ogura, A. Mashiyama, and H. Akai,
 J. Phys. Soc. Jpn. {\bf 84}, 084702 (2015).
 \bibitem{Kou}
{ X. C. Kou, F. R. de Boer, R. Gr\"ossinger, G. Wiesinger, H. Suzuki, H. Kitazawa, T. Takamasu, and G. Kido, J. Magn. Magn. Mater.  {\bf 177--181}, 1002 (1998).
}

%6
\bibitem{XAS}
T. W. Capehart, R. K. Mishra, and F. E. Pinkerton, Appl. Phys. Lett. {\bf 58}, 1395 (1991).

\bibitem{Wijn}
{H. W. de Wijin, A. M. van Diepen, and K. H. J. Buschow, Phys. Rev. B {\bf 7}, 524 (1973).}

\bibitem{Sankar}
{S. G. Sankar, V. U. S. Rao, E. Segal,  W. E. Wallace, W. G. D. Frederick, and H. J. Garrett, Phys. Rev. B {\bf 11}, 435 (1975).}

\bibitem{Yamada}
M. Yamada, H. Kato, H. Yamamoto, and Y. Nakagawa, Phys. Rev. B {\bf 38}, 620 (1988).
\bibitem{Richter}
{ M. Richter J. Phys. D: Appl. Phys. {\bf 31} 1017 (1998).}

\bibitem{Franse}
{ J. Franse and R. Radwa\'{n}ski, in {\it Handbook\ of\ Magnetic\ Materials}, edited by K. H. J. Buschow (North-Holland, Amsterdam, 1993), Vol. 7.}
%7
\bibitem{Yoshiokasan}
T. Yoshioka, and  H. Tsuchiura,  Appl. Phys. Lett. {\bf 112}, 162405 (2018).
\bibitem{Yoshioka1-12}
 T. Yoshioka, H. Tsuchiura, and P. Nov\'{a}k, Phys. Rev. B {\bf 102}, 184410 (2020).
\bibitem{Novak}
P. Nov\'ak, Phys. Stat. Sol. B {\bf 198}, 729 (1996).
%10
\bibitem{Divis1}
M. Divi\v s, K. Schwarz, P. Blaha, G. Hilsher, H. Michor, and S. Khmelevskyi, Phys. Rev. B {\bf 62}, 6774 (2000).
%11
\bibitem{Divis2}
M. Divi\v s, J. Rusz, H. Michor, G. Hilsher, P. Blaha, and K. Schwarz, J. Alloys Compd. {\bf 403}, 29 (2005).
\bibitem{wien2k}
P. Blaha, K. Schwarz, G. K. H. Madsen, D. Kvasnicka, and J. Luitz, {\it WIEN2k,\ An\ Augmented\ Plane\ Wave\ +\ Local\ Orbitals\ Program\ for\ Calculating\ Crystal\ Properties}  (Karlheinz Schwarz, TU Wien, Austria, 2001).%, ISBN 3-9501031-1-2.
%9

\bibitem{Inami}
N.  Inami, Y. Takeuchi, T. Koide, T. Iriyama, M. Yamada and Y. Nakagawa, J. Appl. Phys. {\bf 115}, 17A712 (2014).
%8
\bibitem{Brennan}
S. Brennan, R. Skomski, O. Cugat and J. M. D. Coey, J. Magn. Magn. Mater. {\bf 140--144}, 971 (1995). 
\bibitem{Kuzmin}
M. D. Kuz'min, Phys. Rev. Lett. {\bf 94}, 107204 (2005).
\bibitem{koyama}
 K. Koyama and H. Fujii, Phys. Rev. B {\bf 61}, 9475 (2000).
%12
\bibitem{Sasaki}
R. Sasaki, D. Miura, and A. Sakuma Appl. Phys. Express {\bf 8}, 043004 (2015).
%13

%14
\bibitem{Brooks}
M. S. S. Brooks, L. N\"ordstrom, and B. Johansson, J. Phys. Condens. Matter {\bf 3}, (1991). %{\color{red}きになる}
\bibitem{Teresiak}
A. Teresiak, M. Kubis, N. Mattern, K.-H. Mller, and B. Wolf, J. Alloys Compd. {\bf 319}, 168 (2001).
\bibitem{McClure}
D. S. McClure, Solid State Phys. {\bf 9}, 399--525 (1959). % {\color{red}ページ数確認}

%\bibitem{Kronmuller}
%H. Kronm\"uller, K. D. Drust, and G. Martinek, J. Magn. Magn. Mater. {\bf 69}, 149--157 (1987).
% Kronmuller equation

%\bibitem{Fischbacher}
%J. Fischbacher, A. Kovacs, H. Oezelt, M. Gusenbauer, T. Schrefl, L. Exl, D. Givord, N. M. Dempsey, G. Zimanyi, M. Winklhofer, G. Hrkac, R. Chantrell, N. Sakuma, M. Yano, A. Kato, T. Shoji, and A. Manabe, Appl. Phys. Lett. {\bf 111} 072404 (2017).
% main phase & grain boundary
%
%\bibitem{Fukagawa}
%T. Fukagawa and S. Hirosawa, J. Appl. Phys. {\bf 104}, 013911 (2008).
%
%\bibitem{Amin1}
%H. Sepehri-Amin, T. Ohkubo, T. Nishiguchi, S. Hirosawa, and K. Hono, Scr. Matter. {\bf 63}, 1124 (2010).
%
%\bibitem{Amin2}
%H. Sepehri-Amin, T. Ohkubo, S. Nagashima, M. Yano, T. Shoji, A. Kato, T. Schrefl, K. Hono, Acta. Mater. {\bf 61}, 6622--6634 (2013).
%\bibitem{hono}
 %K. Hono and H. Sepehri-Amin, Scr. Matter. {\bf 67}, 530 (2012).
% surface effeccts
%\bibitem{Moriya}
% H. Moriya, H. Tsuchiura, and A. Sakuma, J. Appl. Phys. {\bf 105} 07A740 (2009).
%
%\bibitem{Tanaka}
% S. Tanaka, H. Moriya, H. Tsuchiura, A. Sakuma, M. Divi\v s, and P. No\'ak, 
% J. Appl. Phys. {\bf 109}, 07A702 (2011).
%
%\bibitem{tsuchi}
% H. Tsuchiura, T. Yoshioka, and P. Nov\'{a}k, IEEE Trans. Magn. {\bf 50}, 2105004 (2014).
% Atomistic LLG
%\bibitem{Mitsumata}
%C. Mitsumata, H. Tsuchiura, and A. Sakuma, Appl. Phys. Express {\bf 4}, 113002 (2011).
%
%\bibitem{Hummler1}
%K. Hummler and M. F\"ahnle, Phys. Rev. B {\bf 53}, 3290 (1996).
%
% method
%\bibitem{Stevens}
% K. Stevens, Proc. Phys. Soc. A {\bf 65}, 209 (1952).
%
%\bibitem{Hutchings}
 %M. T. Hutchings, Solid State Phys. {\bf 16}, 227 (1964).
%
%\bibitem{Richter}
%M. Richter, J. Phys. D {\bf 31}, 1017 (1998).
%
% open core
\bibitem{Pavel_Wannier}
 P. Nov\'{a}k, K. Kn\'{i}\v{z}ek, and J. Kune\v{s}, Phys. Rev. B {\bf 87}, 205139 (2013). 
\bibitem{Wirth}
{ S. Wirth, M. Wolf, K.-H. M\"uller, R. Skomski, S. Brennan, and J. M. D. Coey, IEEE Trans. Magn. {\bf 32}, 4746 (1996).}
\bibitem{Isnard}
{O. Isnard, S. Miraglia, M. Guillot, and D. Fruchart, J. Appl. Phys.  {\bf 75}, 5988 (1994).}
%Wannierization

%
\bibitem{yoshi_Wannier}
 T. Yoshioka, H. Tsuchiura, and P. Nov\'{a}k, Mater. Res. Innov. {\bf 19}, S3 (2015).
%
% wien2k

% review
%\bibitem{Herbst1}
%J. F. Herbst, Rev. Mod. Phys. {\bf 63} 819 (1991).
%
% the parameters in the spin model
%\bibitem{miura}
% Y. Miura, H. Tsuchiura, and  T. Yoshioka, 
 %J. Appl. Phys. {\bf 115}, 17A765 (2014).
% exp. fitting
%
%\bibitem{Hirosawa} %Ho, Er, Tm
%S. Hirosawa, Y. Matsuura, H. Yamamoto, S. Fujimura, M. Sagawa, and H. Yamaguchi, J. Appl. Phys. {\bf 59} 873 (1986).
% fitting for JFeFe
%\bibitem{Fuerst}
%C. D. Fuerst, J. F. Herbst, and E. A. Alson, J. Magn. Magn. Mater. {\bf 54--57}, 567 (1986).
% anisotropy constant

%\bibitem{Amin3}
 %H. Sepehri-Amin, T. Ohkubo, S. Nagashima, M. Yano, T. Shoji, A. Kato, T. Schre, and K. Hono, 
% Acta. Mater. {\bf 61}, 6622 (2013).
% interfacial structure
%\bibitem{tatetsu}
% Y. Tatetsu, S. Tsuneyuki, and Y. Gohda,
% Phys. Rev. Applied {\bf 6}, 064029 (2016).
%
\end{thebibliography}

%
%%%%%%%%%%%%%%%%%
\end{document}